# Raman properties of GaSb nanoparticles embedded in SiO$_2$ films

Fa-Min Liu [1, 2] *   Tian-Min Wang [1]   Jian-Qi Li [2]   Li-De Zhang [3]   Guo-Hua Li [4]

[1] * Center of Material Physics and Chemistry, School of Science, Beijing University of Aeronautics and Astronautics, Beijing 100083, P. R. China
< Tel. +86-10-82317941, Fax. +86-10-82315933, E-mail: fmliu@aphy.iphy.ac.cn ; fmliu@etang.com >
[2] National laboratory for Superconductor, Institute of Physics, Chinese Academy of Sciences, Beijing100080, P. R. China
[3] Institute of Solid State Physics, Chinese Academy of Sciences, Hefei 230031, P. R. China
[4] National Laboratory for Superlattices and Microstructure, Institute of Semiconductors, Chinese Academy of Sciences, Beijing, 100083, P. R. China

**Abstract:** The Raman shifts of nanocrystalline GaSb excited by an Ar$^+$ ion laser of wavelengths 514.5, 496.5, 488.0, 476.5, and 457.9 nm are studied by experiment and explained by phonon confinement, tensile stress, resonance Raman scattering and quantum size effects. The Stokes and anti-Stokes Raman spectra of GaSb nanocrystals strongly support the Raman feature of GaSb nanocrystals. Calculated optical spectra compare well with experimental data on Raman scattering GaSb nanocrystals.

# 1. Introduction

The behaviour of semiconductor nanostructures embedded in dielectric matrix is significantly different from other low-dimensional semiconductor heterostructures [1-6]. Indeed, electronic states of these nanostructures have to be treated with an account for interactions with real surfaces or interfaces and the surrounding material. Phonon states are also modified in semiconductor nanostructures from those of the bulk. The purpose of the present article is to discuss the Raman properties of GaSb nanocrystals grown in $SiO_2$ host by radio frequency magnetron co-sputtering. We have demonstrated and tentatively explained the shift of the Raman peaks under 514.5, 496.5, 488.0, 476.5, and 457.9 nm excitation lines of an $Ar^+$ ion laser. The Stokes and anti-Stokes Raman spectra of the GaSb nanoparticles embedded in a $SiO_2$ matrix was also carried out for comparison and better understanding of the origin of phonon-confinement and tensile-stress effects.

# 2. Experimental Procedure

GaSb nanocrystals is grown in $SiO_2$ host by radio frequency magnetron co-sputtering, which is described elsewhere [7]. The target used was a $SiO_2$ glassy plate 60 mm in diameter, with several chips of polycrystalline GaSb attached. By varying the numbers of GaSb chips, the semiconductor concentration in the deposited film can be controlled to a desired value. In this work the effective sputtered area ratio of the semiconductor chips to $SiO_2$ plate was fixed at 20%. Some glass substrates (10 mm × 10 mm ×1mm) were ultrasonically washed successively in acetone, alcohol, and de-ionized water to obtain a clean surface before being placed in a vacuum chamber. The target was separated from the substrate by 5 cm. When the chamber was evacuated to a pressure of $2\times10^{-5}$ Torr high purity (99.9999%) argon gas was introduced. During sputtering, the chamber pressure was maintained at $1.5 \times 10^{-2}$ Torr. The output voltage to the radio frequency sputter gun was 1000 V. All runs started with presputtering the target for 15 minutes to ensure good quality GaSb films. The films were grown at various substrate temperatures and with various duration.

Room temperature Raman spectra of GaSb nanocrystals were recorded with an SPEX-1403 laser Raman spectrometer. The excitation source was an $Ar^+$ ion laser of 514.5, 496.5, 488.0, 476.5, and 457.9 nm lines operated in the backscattering geometry and the resolution of the instrument was 1 $cm^{-1}$. The Stokes and anti-Stokes Raman spectra of GaSb nanocrystals were also recorded with a SPEX-1403 laser Raman spectrometer with excitation of an argon ion laser of 514.5 nm at room temperature in the back-scattering configuration. The output power of the laser was kept within 200 mW and a cylindrical lens was used to avoid overheating of the samples.

3. Results and Discussion

The nanocrystalline $GaSb-SiO_2$ films were characterized with a rotating anode D／max−rA type X-ray diffractometer with an operating voltage of 40 kV and a current 100 mA. The experimental result [7] indicates that the GaSb nanocrystals are a typical zinc-blende structure. The average size of the nanocrystals is about 4.1 nm estimated by the X-ray diffraction peak using the Scherrer method.

Fig.1 shows the Raman shifts of nanocrystalline GaSb when under excitation by lasers of wavelengths 514.5, 496.5, 488.0, 476.5, and 457.9 nm. From the Fig.1, we can see red shifts of Raman peaks as the excited wavelength increases. The shifts of Raman peaks excited by the different wavelength are shown Table 1. One can see that when the excitation lines are from 457.9 to 514.5 nm, the Raman peaks of nanocrystalline GaSb have 15.9 cm$^{-1}$ red shifts, and the full width at half maximum (FWHM) of Raman peak is widened about 7.5 cm$^{-1}$. We have also found that the Fig.1 and Table 1 are independent of the intensity of excitation line. The Raman spectra of nanocrystalline GaSb with respect to that of the bulk GaSb was published in elsewhere [7]. Compared to that of bulk GaSb [7,11], which the TO and LO modes are at 225.6 and 236.1 cm$^{-1}$, respectively, there is a large red shifts of Raman peaks of nanocrystalline GaSb originated from phonon confinement [7-9,11] and tensile stress effects [10-11].

First, we discuss the Raman peak red shift by the phonon confinement effect using the theory of Richter [8] and Campbell [9] et al. Richter et al. [8] considered only spherical microcrystals and compared their results obtained from silicon films prepared by the plasma transport method. Campbell et al. [9] further considered the exact effects of different microcrystal sizes and shapes. They showed that there are significant differences between spherical, columnar, and thin slab microcrystals. They neglected scale factors and assumed the phonon mode at the reduced wavevector starting point $q_o$ = 0 which is appropriate for one phonon scattering. The first order Raman spectrum, I ($\omega$), is

$$I(\omega) \cong \int d^3q \, |C(0,q)|^2 / \{[\omega-\omega(q)]^2 + (\Gamma_o/2)^2\} \qquad (1)$$

Where $I(\omega)$ represents the confined Raman line shape, $d^3q$ is the volume in momentum space, $\omega(q)$ is the phonon dispersion curve, $\Gamma_o$ is the natural linewidth, and $C(0, q)$ is the Fourier coefficient of the phonon confinement function. From formula (1), the central frequency red shift of nanocrystalline $GaSb$-$SiO_2$ composite films is estimated about several $cm^{-1}$. Obviously, it is smaller than that of our experimental results. So, another source of red shift of the Raman peak should be considered. Second, in order to explain the results, we should further consider the tensile stress caused by the mismatch of the different thermal expansion coefficients. The hint figure of GaSb nanocrystals embedded in $SiO_2$ host is shown in Fig.2. One can obviously see that there are a large number of atoms on the surface of GaSb nanocrystals due to quantum size effect. And also, it is well known that there are three forces ($F_1$, $F_2$ and $F_3$ shown in Fig. 2) exerting on the nanocrystalline GaSb. $F_1$, called for interface covalent force, is originated from Sb=O owing to the binding energy of Sb3d closing to that of O1s. The XPS and XRD data [7,11] strongly support the tensile stress existed in the GaSb nanocrystals. $F_2$ is called for thermal expansion force caused by the mismatch of the different thermal expansion coefficients, and $F_3$ is considered as the compressed force of $SiO_2$ matrix. The resultant force F exerted on the GaSb nanocrystals is equal to $F_1 + F_2 - F_3$. The average stress is about $1.6 \times 10^{10}$ N/$m^2$ estimated according to the XRD of the composite film [7,12]. Richter et al. [8] pointed out that tensile and compressive stress affect the Raman line by a redshift and a blueshift respectively. Raisin et al. [10] studied

GaSb/GaAs heteroepitaxy by Raman spectroscopy. The value of the residual strain ε can be deduced from the LO single relative frequency shift by [10]

$$\Delta\omega/\omega=[(S_{11}+2S_{12})(K_{11}+2K_{12})-(S_{11}-S_{12})(K_{11}-K_{12})]\varepsilon/6(S_{11}+S_{12}) \qquad (2)$$

Where the $S_{ij}$ are the elasticity constants and $K_{ij}$ the reduced phonon deformation potentials.

Therefore, as an example, we have estimated the Raman shift of the LO mode. From formula (1) the Raman shift was estimated to be about 6.21 cm$^{-1}$. Also, from formula (2) the Raman shift was estimated to be about 86.11 cm$^{-1}$. These are basically consistent with the experimental results. Consequently, the large Raman red shift of the GaSb-SiO$_2$ composite film compared to that of the bulk GaSb could be explained by tensile stress and phonon confinement effects.

The Raman phenomena mentioned above is also found in porous Si [13,14], silicon nanowires [15,16], silicon nanocrystals [17], and carbon nanotube [18,19]. Zhang, et al. [13] studied the intrinsic Raman spectrum of porous silicon formed on non-degenerate p-type silicon by using the different incident laser wavelength. They found that intrinsic Raman frequencies have small up-shift variations, whereas the Raman linewidths becomes a narrow line, with increasing the wavelength of the lasers. These phenomena were interpreted by using the phonon confinement effect and resonance Raman scattering. Rao, et al. [18] studied diameter selective Raman scattering from vibrational modes in carbon nanotubes. They found resonant Raman scattering process with laser excitation wavelengths in the range from 514.5 nm to 1320 nm. This resonance results from one-dimensional quantum confinement of the electrons in the nanotube. Milnera, et al. [19] studied the resonance excitation and intertube interaction from quasicontinuous distributed helicities in single wall carbon nanotubes. They used a large number of different lasers extending from the deep blue (459 nm) to the infrared (1068.6 nm). They found that the resonance excitation exhibited an oscillatory behavior and had up-shifted by 8.5% on the average. Obviously, the experimental results of nanocrystalline GaSb described in Fig.1 are contrary to that of porous silicon, silicon nanowires and nanotubes. Firstly, the Raman scattering has blue shifts (up-shifts) with increasing the excitation wavelength of lasers for silicon nanocrystals and nanotubes, whereas the Raman scattering has red shifts (down-shifts) for nanocrystalline GaSb with increasing the excitation wavelength of lasers. Secondly, the Raman peaks of silicon nanocrystal and nanotubes mainly scatter from Si-Si and C-C bonds respectively, whereas the Raman peaks of GaSb nanocrystals mainly scatter from Ga-Ga, Ga-Sb and Sb-Sb bonds. Moreover, GaSb nanocrystals were affected by tensile stress because it was embedded in $SiO_2$ matrix. So we have to find another view to interpret our experimental results. We think that there is a distribution of GaSb nanoparticles in the films, such as a Lorentzian or Gaussian size distribution function. Owing to the quantum size effect, the GaSb nanocrystals are vibrating when excited by a laser. So, we can consider the GaSb nanoparticles as much optical vibrons [20] which has different momentum. Trallero-Giner, et al. [20] studied the optical vibrons in CdSe quantum dots. They presented a detailed analysis of Raman line shape in terms of size distribution, phonon linewidths, and photon energy of the exciting laser. They founded that the relative contribution of the excitonic and vibronic state to the Raman line shape depends on the incoming laser frequency, the excitonic oscillator strength, and the exciton-vibron interaction. Raman scattering can be used to describe a interaction process of photon-electron-phonon. Saio, et al. [21] studied the phonon dispersion relation by double resonance Raman scattering. They thought that in the Stokes (phonon emission) process, an electron with momentum k [or electron energy $E^i(k)$] is excited by the incident laser photon in an electron-hole creation process. The electron is then scattered by emitting a phonon with momentum q to the state with momentum k + q [or energy E(k + q)], then scattered again, back to k [$E^f(k)$] to recombine with a hole. In the one-phonon emission Stokes process, one of the two scattering processes ( k → k + q, k + q → k) is inelastic with a phonon emission process, and the other is an elastic scattering process mediated by the effect. So we think that these nano-vibrons, which have the same momentum as that of the photons, will be got a prior energy to vibrate. Therefore, when excited by short wavelength light, the nano-vibrons will be got very high frequency vibrating and scattering because the photon has more high momentum, so the blue Raman shifts (up-shifts) will be measured. Whereas, when excited by the long wavelength light, the nano-vibrons will be got very low frequency vibrating and scattering because the photon has less momentum, so the red Raman shifts (down-shifts) will be measured. In addition, there are many involved factor such, as the inner stress in the film, the shapes of the GaSb nanocrystals, will bring a very small Raman shifts.

The Stokes and anti-Stokes Raman spectra of GaSb nanocrystals are shown in Fig.3, which is excited by a 514.5 nm line of $Ar^+$ laser. One can see the enhanced Stokes and anti-Stokes Raman peaks of nanocrystalline GaSb compared with that of bulk GaSb. And, the intensity of Stokes Raman peak is larger than that of anti-Stokes. The relative intensities $I_S / I_{AS}$ for a given laser line should be approximately equal to $exp(\omega_0 /kT)$, where $\hbar\omega_0$ is the phonon energy [22]. At room temperature, $kT' \approx 24$ meV and the optical-phonon energies lie in the range 25-65 meV so that the Stokes line can be much stronger than the anti-Stokes one. This demonstrated that the Raman scattering phenomena of nanocrystalline GaSb described in Fig.1 is truly Raman active modes. A large red shifts of Raman peaks and an enhanced Raman intensity is originated from the phonon confinement [8,9] and tensile stress effects [10]. But, another anomalous Raman scattering phenomena of nanocrystalline GaSb, which is not explained by the Raman selection rules, is that the absolute value of the Stokes Raman shifts is little more large (about 1.8 $cm^{-1}$) than that of the anti-Stokes. This is originated from the reduced symmetry for nanocrystals and quantum wires [23]. So we think that there are the size distributions of GaSb nanocrystals in the $GaSb-SiO_2$ composite films, and the different Raman shifts is got for different size nanoparticles of GaSb owing to the quantum size effect. From the theory of quantum statistics, one knows the atomic number of the ground state is very larger than that of the excited state. Therefore, the iterative results of Raman scattering of different nanopaticles are that the Stokes Raman shifts are little larger than that of the anti-Stokes. According to the feature of the Stokes and anti-Stokes, one can estimate the sample temperature from the intensity ratio between Stokes and anti-Stokes Raman modes according to the relation [24]

$$I_{Stokes}/I_{anti-Stokes} = [(\omega_L-\omega)/(\omega_L+\omega)]^4 \exp(2h\omega/k_B T) \quad (3)$$

Where $\omega_L$ and $\omega$ are the frequency of incident laser light and the phonon of the Raman mode considered, respectively, $k_B$ is Boltzmann's constant, and T is the sample temperature. In our experiment, the sample is kept at room temperature (~300K) because no frequency shift induced by the change of the sample temperature is observed when the excitation power is much lower than the current laser density.

## 4. Conclusion

In summary, we have successfully prepared the films of nanocrystalline GaSb embedded in the $SiO_2$ host by radio frequency magnetron co-sputtering. The Raman shifts of nanocrystalline GaSb excited by an $Ar^+$ ion laser of 514.5, 496.5, 488.0, 476.5, and 457.9 nm lines are studied by experiment and explained by phonon confinement, tensile stress, resonance Raman scattering and quantum size effects. It is shown that there is a small change of Raman shifts for different excitation lines. We have shown that the measurement of Stokes and anti-Stokes Raman spectra of GaSb nanocrystals excited by 514.5 nm lines of an argon laser. These results strongly support the Raman feature of GaSb nanocrystals. The estimated Raman shifts of GaSb nanocrystals compare well with experimental data according to phonon confinement and tensile stress effects.

**Acknowledgments**

This work was supported by The National Climbing Program of China and in part by The Postdoctoral Foundation of China.

**Reference**

[1] Yoshida S, Hanada T, Tanabe S, and Soga N, 1999 J. Mater. Sci. **34** 267.

[2] Chen W, Cai W P, Liang, C H and Zhang L D, 2001 Mater. Res. Bull. **36** 335.

[3] Gangopadhyay P, Kesavamoorthy R, Nair K G M, and Dhandapani R, 2000 J. Appl. Phys., **88** 4975.

[4] Li G H, Ding K, Chen Y, Han H X, and Wang Z P, 2000 J. Appl. Phys., **88** 1439.

[5] Kolobov A V, Maeda Y, and Tanaka K, 2000 J. Appl. Phys., **88** 3285.

[6] Arora A K, and Rajalakshmi M, 2000 J. Appl. Phys., **88** 5653.

[7] Liu F M, and Zhang L D, 1999 J. Crystal Growth, **204** 19.

[8] Richter H, Wang Z P, and Ley L, 1981 Solid State Commun. **39** 625.

[9] Campbell I H, and Fauchet P M, 1986 Solid State Commun. **58** 739.

[10] Raisin C, Rocher A, Landa G, Cares R, and Lassabatere L, 1991 Appl. Sur. Sci. **50** 434.

[11] Liu F M, and Zhang L D, 1999 Semicond. Sci. Tech., **14** 710.

[12] Liu F M, Zhang L D, Zheng M J, and Li G H, 2000 Appl. Surf. Sci., **158** 281.

[13] Zhang S L, Hou Y T, Ho K S, Qian B D, and Cai S M, 1992 J. Appl. Phys. **72** 4469.

[14] Zhang S L, Wang X, Ho K S, Li J J, Diao P, and Cai S G, 1994 J.Appl.Phys. **76** 3016.

[15] Yu Y D, Bai Z G, Feng S Q, Lee C S, Bello I, Sun X S, Tang Y H, Zhou G W, Zhang Z, 1998 Solid State Commu., **105** 403.

[16] Li B B, Yu D P, and Zhang S L, 1999 Phys. Rev. B **59** 1645.

[17] Sirenko A A, Fox J R, Akimov I A, Xi X X, Ruvimov S, and Liliental-Weber Z, 2000 Solid State Commu. **113** 553.

[18] Rao A M, Richer E, Bandow S, Chase B, Eklund P C, Williams K A, Fang S, Subbaswamy K R, Menon M, Thess A, Smalley R E, Dresselhaus G, and Dresselhaus M S, 1997 Science, **275** 187.

[19] Milnera M, Kurti J, Hulman M, Kuzmany H, 2000 Phys. Rev. Lett. **84** 1324.

[20] Trallero-Giner C, Debernardi A, Cardona M, Menendez-Proupin E, Ekimov A I, 1998 Phys. Rev. B, **57** 4664.

[21] Saito R, Jorio A, Souza Filho A G, Dresselhaus G, Dresselhaus M S, Pimenta M A, 2002 Phys. Rev. Lett. **88** 027401.

[22] Trzeciakowski W, Martinez-Pastor J, and Cantarero A, 1997 J.Appl.Phys., **82** 3976.

[23] Mariani E, Sassetti M, Kramer B, 2000 Europhys. Lett., **49** 224.

[24] Hayes W, Loudon R, The Scattering of Light by Crystals (Wiley, New York, 1976).

**Captions for Figures**

Fig.1 The Raman shifts of nanocrystalline GaSb excited by an $Ar^+$ ion laser of the different lines

Table 1  The shifts of Raman peaks excited by the different wavelength

Fig. 2  The hint figure of GaSb nanocrystals embedded in $SiO_2$ host

Fig.3 The Stokes and anti-Stokes Raman spectra of GaSb nanocrystals

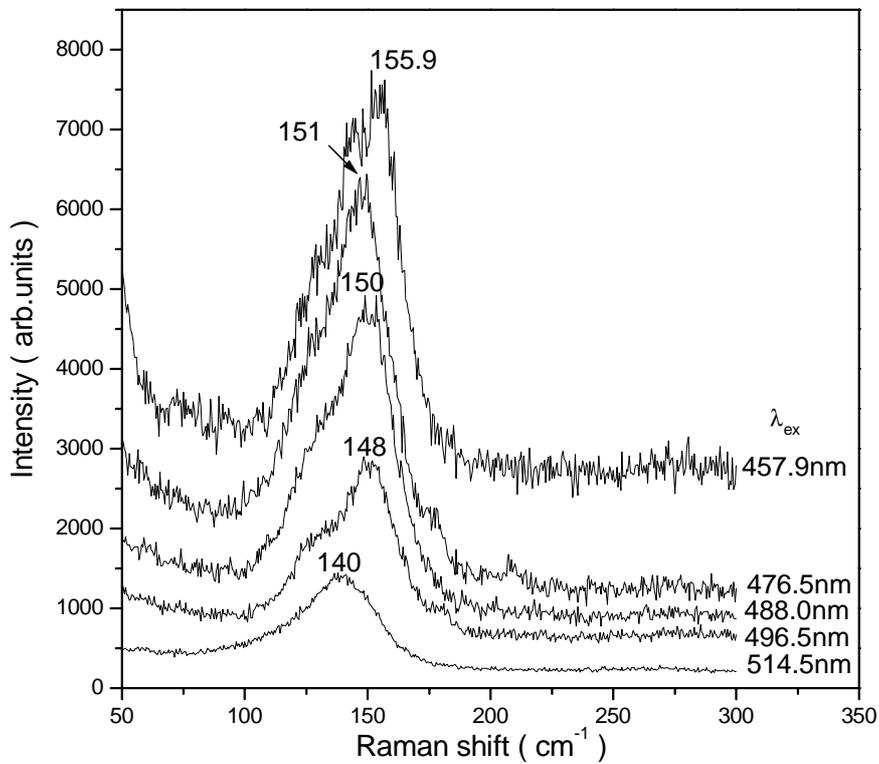

Fig.1 The Raman shifts of nanocrystalline GaSb excited by an Ar$^+$ ion laser of the different lines

By F.M. Liu, et al.

Table 1 The shifts of Raman peaks excited by the different wavelength

| Wavelength of the excitation lines ( nm ) | Position of the Raman peak ( cm$^{-1}$ ) | Shifts of the Raman peak ( cm$^{-1}$ ) | Variations of full width at half maximum ( FWHM ) of Raman peak ( cm$^{-1}$ ) |
|---|---|---|---|
| 514.5 | 140.0 |  | 46.2 |
| 496.5 | 148.0 | 8 | 45.2 |
| 488.0 | 150.0 | 10 | 44.1 |
| 476.5 | 151.0 | 11 | 43.2 |
| 457.9 | 155.9 | 15.9 | 38.7 |

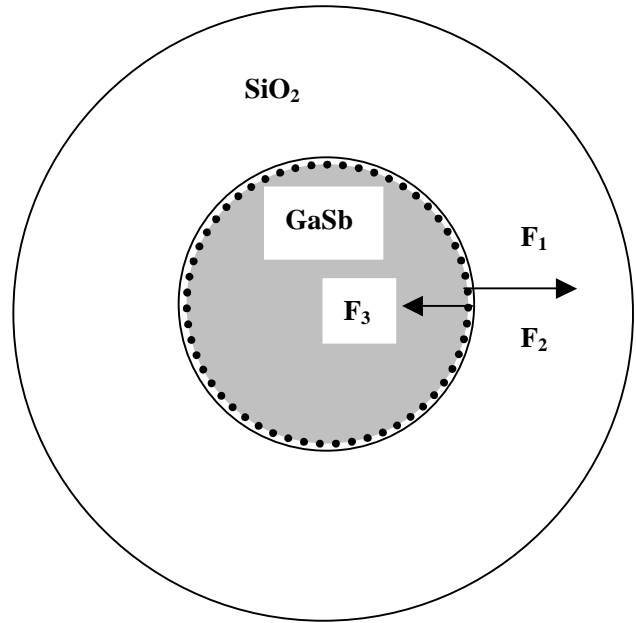

Fig. 2　The hint figure of GaSb nanocrystals embedded in SiO$_2$ host

By F. M. Liu, et al.

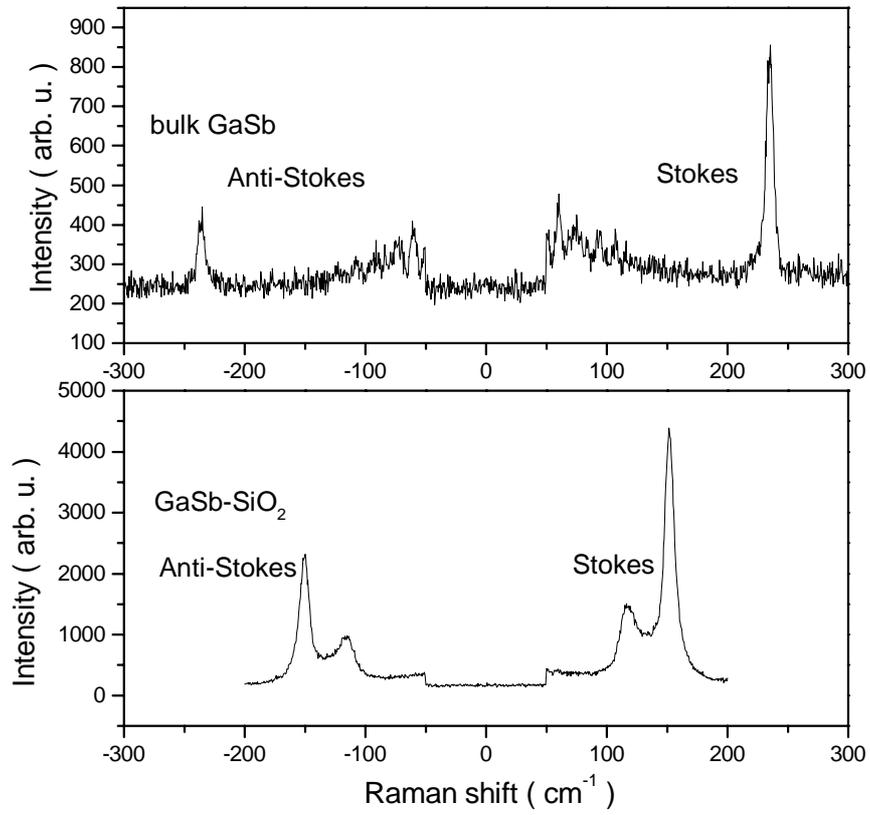

Fig. 3  The Stokes and Anti-Stokes Raman spectra of GaSb nanocrystals

By F. M. Liu, et al.